\documentclass[superscriptaddress,footnoteinbib,showpacs,twocolumn,amssymb,aps]{revtex4-1}

\usepackage{graphicx,amsmath,color}

\newcommand{\fig}[1]{Fig.~\ref{#1}}
\newcommand{\eq}[1]{Eq.~(\ref{#1})}

\newcommand{\eeq}{ \end{equation} }
\newcommand{\beq}{ \begin{equation} }
\newcommand{\eea}{ \end{eqnarray} }
\newcommand{\bea}{ \begin{eqnarray} }

\newcommand{\bhu}{ {\bf \hat{u}} }
\newcommand{\bhe}{ {\bf \hat{e}} }

\newcommand{\bxi}{ {\bf \xi}}
\newcommand{\bhz}{ {\bf \hat{z}} }

\newcommand{\bff}{ {\bf f }}

\newcommand{\br}{ {\bf r }}

\newcommand{\bw}{ {\bf w }}

\newcommand{\Dbar}{ \overline{D}}

\newcommand{\bfr}{ {\bf r} }

\newcommand{\bfq}{ {\bf q} }
\newcommand{\bfnp}{ {\bf n}_{\perp} }
\newcommand{\bfqp}{ {\bf q}_{\perp} }
\newcommand{\bfrp}{ {\bf r}_{\perp} }
\newcommand{\bzh}{ {\bf \hat{z}} }
\newcommand{\bn}{ {\bf \hat{n}} }

\newcommand{\kbt}{k_{\rm B}T}

\begin{document}

\title{Long-time anomalous swimmer diffusion in smectic liquid crystals}
\author{Claudia Ferreiro-C\'{o}rdova}
\affiliation{Laboratoire de Physique des Solides, Universit\'e Paris-Sud  \& CNRS, UMR 8502, 91405 Orsay, France}
\author{John Toner}
\affiliation{Department of Physics and Institute of Theoretical Science, University of Oregon, Eugene, OR 97403, USA}
\author{Hartmut L\"{o}wen}
\affiliation{Institut f\"{u}r Theoretische Physik II: Weiche Materie,
Heinrich-Heine-Universit\"{a}t D\"{u}sseldorf, Universit\"{a}tsstra{\ss}e 1, 40225 D\"{u}sseldorf, Germany}
\author{Henricus H. Wensink }
\affiliation{Laboratoire de Physique des Solides, Universit\'e Paris-Sud  \& CNRS, UMR 8502, 91405 Orsay, France}
\email{rik.wensink@u-psud.fr}

\date{\today}

\begin{abstract}
The dynamics of self-locomotion of active particles in aligned or  liquid crystalline fluids strongly deviates from that in simple isotropic media.    We explore the long-time dynamics of a swimmer moving in a  three-dimensional smectic liquid crystal and find that the mean-square displacement (MSD) transverse to the director exhibits a distinct logarithmic tail at long times. The scaling is distinctly different from that  in an isotropic  or nematic fluid and hints at the subtle but important role of the director fluctuation spectrum in governing the long-time motility of active particles. Our findings are based on a generic hydrodynamic theory and Brownian dynamics computer simulation of a three-dimensional soft mesogen model. 

\end{abstract}

\maketitle

\section{Introduction}
The main focus of research in the active matter physics community has recently shifted towards studying microswimming through complex media that can no longer be re-presented by a simple isotropic Newtonian continuum \cite{Marcetti_RMP,romanczuk2012,Gompper_Winkler_review,bechinger_volpe2016,hagan_baskaran2016}.
The main motivation stems from the fact that many microorganisms operate in crowded
environments with a non-uniform positional and/or orientational microstructure that may generate liquid crystalline or viscoelastic properties. Examples encompass the dynamics of cilia and spermatozoa
in mucus \cite{ref4,ref7}, of bacteria migrating through 
tissue \cite{ref6}, or through complex extracellular matrices such as in biofilms and the motion of nematodes residing in  soil \cite{ref5}.

These real-life situations call for more sophisticated models for microswimming that aim at a better understanding of the complexity of the medium, such as in the case of viscoelastic
fluids \cite{FuPowers2007,Arratia1,Arratia2,Liu,Poon_PNAS,Peer_Fischer,Lauga0,Lauga1,Lauga2,bechinger2016,bechinger2018}, 
or liquid crystalline fluids 
\cite{shi-powers2017,lintuvuori_wurger2017,toner_wensink2016, PNAS_Aranson,Lavrentovich2015,Lauga2,Mushenheim,Sagues,Krieger1,Krieger2,trivedi2015}. A similar increase in medium complexity is attained by considering active locomotion around random or patterned obstacles \cite{peruaniPRL2013,volpeSM2011,Morin2017}, and by studying the role of active dopants in crystalline host systems \cite{Krieger3,Filion,Poon2016}.

With most approaches thus far focussing on collective properties or on short-term swimmer motility, we wish to address the impact of  
liquid crystalline order on the long-time diffusive behavior of such a swimmer. 
In a previous paper  \cite{toner_wensink2016}, we have undertaken such a study  by focussing on active diffusion through a simple Lebwohl-Lasher lattice nematic. Here, we wish to build upon these findings and consider a more appropriate {\em off-lattice} model to explore long-time active diffusion in lamellar or smectic systems which possess a distinct unidimensional long-range periodicity imparted by `stacked' membranes each with a quasi-bidimensional liquid-like internal order. In contrast to the simplified lattice representation of a liquid crystal, the off-lattice model enables us to vary the positional symmetry of the host medium simply by changing the system temperature.  The model thus offers a route to sampling the non-trivial swimmer dynamics moving through a range of different host phases and use temperature as a control parameter. We find that in the case of nematic and smectic hosts, the swimmer MSD perpendicular to the director is non-trivial and obeys a distinct logarithmic scaling with time. Specifically, we find that the mean squared lateral wandering $\left<\left( \Delta \bfr_{s}^{\perp}(t)\right)^{2}\right>$ of such a swimmer obeys
\begin{eqnarray}
\left<\left( \Delta \bfr_{s}^{\perp}(t)\right)^{2}\right> =\left\{
\begin{array}{ll}
 \overline{D}_s t\ln\left({t\over t_0}\right), \qquad
&t\ll t_a\,
\\ \\
 \overline{D}_{s} t\sqrt{\ln\left({t\over t_0}\right)}, \qquad
&t\gg t_a\,
\end{array}\right.
\label{sm_scaling}
\end{eqnarray}
where $\overline{D}_{s}$ represents an anomalous diffusion constant and $t_a$ is a system-dependent crossover time that is extremely sensitive to both the speed of the swimmer and the parameters of the smectic liquid crystal. An explicit expression for $t_a$ is given in equations (\ref{tacon}) and  (\ref{vcdef}).

In contrast, it was shown in  \cite{toner_wensink2016} that a swimmer in a nematic also exhibits anomalous diffusion, but obeys  the law
\begin{eqnarray}
\left<\left( \Delta \bfr_{s}^{\perp}(t)\right)^{2}\right> = \overline{D}_n t\ln\left({t\over t_0}\right) \,\,\,\,\,\,\,\,\,.
\label{nem_scaling}
\end{eqnarray}
Since the typical scaling  exponent for a smectic fluid differs from that of a nematic system, measurement of the swimmer MSD transverse to the main director of the host  could be used to probe structural features of the medium, in particular the presence of local lamellar order.   
Of course, to observe this difference, experiments must probe times $t\gg t_a$, since the scaling for shorter times is the same in both the smectic and nematic phases. In many experimental systems, this will be quite difficult: because of the exponential dependence of the crossover time $t_a$ (\ref{tacon}) on material parameters and swimming speed, this time will be literally astronomical in many cases (see the estimates  after equation (\ref{vcdef}) below). The best hope of seeing the $t\gg t_a$ limit of (\ref{sm_scaling}) is for very fast swimmers ($v\sim 50 {\mu\over\rm{sec}}$) in very high dilution lyotropic smectics (see the estimate (\ref{vclamnum}) below for such smectics). 

In our simulations, we circumvent this difficulty by simulating non-momentum conserving dynamics, which, although unphysical for real experiments, reduce the crossover time to extremely small values. The asymptotic form of (\ref{sm_scaling}) for $t\gg t_a$ is unchanged by this change in the dynamics; it is simply reached at much shorter times. And, indeed, these simulations agree with
our theoretical predictions which underscores the fundamental impact of director fluctuations in steering swimmers through liquid crystalline backgrounds. 

The rest of this paper is structured as follows. In Section II we briefly recapitulate our hydrodynamic theory of swimmer motility in anisotropic media and provide new results for the long-time swimmer dynamics in a three-dimensional smectic phase. These predictions are generic and are valid for swimmers in both thermotropic and lyotropic smectic hosts.   The predictions are tested against Brownian dynamics computer simulations based on a soft-nematogen model which is described in detail in Section III. In the  last Section, we formulate the main conclusions of our study. 

\section{Model for long-time swimmer motility} 

In very close analogy with earlier treatments of swimmers in nematics \cite{toner_wensink2016}, we will consider a  self-propelled swimmer moving through an otherwise equilibrium, ordered  smectic A. This swimmer has no memory, or, at best, only a short term memory,  of its past direction of motion. Furthermore, the dynamics of the entire system (smectic plus swimmer) are rotation invariant: that is, the swimmer  carries no internal ``compass"; any preference it exhibits for one direction of motion over any other must arise from the {\it local} layer normal $\bn(\bfr_s(t))$ at the current location $\bfr_s(t)$ of the swimmer. This requirement of locality arises from the physically reasonable assumption that the interactions of the swimmer with the surrounding smectic are short-ranged in space. 
 
The average value of the  instantaneous  velocity $ d \bfr_{s}(t) / dt$ of such a swimmer {\it must} be along $\bn(\bfr_s(t))$; rotation invariance plus locality allow no other direction (except $-\bn(\bfr_s(t))$; we will discuss this option below). Hence, the instantaneous velocity $ d \bfr_{s}(t) / dt $ must be given by
\beq
{d \bfr_{s}(t)\over dt}=v_s\bn(\bfr_{s}(t),t)+\bff(t) \,,
\label{align}
\eeq
where $\bff(t)$ is a zero mean random fluctuation in the  velocity, and $v_s$ is the mean speed of the swimmer. Note that in general $v_s\ne v^0_s$, where  $v^0_s$ is the ``bare", or instantaneous, speed of the swimmer, due to the effects of fluctuations. Indeed, in general, we expect $v_s < v^0_s$. In practice, $v_s$ can only be determined by measuring the mean motion of the swimmer over long times; this will be discussed in more detail below.

The statistics of the fluctuations $\bff$ are also almost completely determined by the requirements of rotation invariance and locality in space and time. In a ``coarse-grained" theory, in which we imagine having averaged our dynamics over time scales long compared to the time of individual molecular ``kicks" experienced by the swimmer, but short compared to the time scales we wish to investigate, $\bff$ can be thought of as a sum of a large number of random molecular kicks at different microscopic times, which are therefore statistically independent. The central limit theorem then tells us that the statistics of $\bff$ should be Gaussian. Its statistics are then completely specified by its two point correlations with the local layer normal $\bn(\bfr,t)$ and itself; rotation invariance and spatio-temporal locality  imply
that these are given by:
\begin{eqnarray}
  \langle f_{\alpha}(t)f_{\beta}(t')\rangle &= 2\Delta_{I}
\delta_{\alpha\beta}\delta(t-t') +2\Delta_{A} n_{\alpha}(\bfr_{s}(t),t) \nonumber \\ 
&\times  n_{\beta}(\bfr_{s}(t),t)   \delta(t-t') \,,
\label{ffcorr}
\end{eqnarray}
and
\begin{eqnarray}
   \langle f_{\alpha}(t)n_{\beta}(\bfr_{s}(t'),t')\rangle=2\Delta_{fn}
\delta_{\alpha\beta}\delta(t-t')\,,
\label{fncorr}
\end{eqnarray}
where $\alpha$ and $\beta$ are Cartesian indices, and $\Delta_{I}$, $\Delta_{A}$, and $\Delta_{fn}$ are phenomenological parameters which set the size of the fluctuations of the swimmer. Because the swimmer is a non-equilibrium agent, these parameters do not, in general, satisfy any kind of Einstein relation; that is, they are independent parameters.

The model just described neglects ``hairpin turns":  fluctuations in which the swimmer reverses its direction of motion relative to the local layer normal (that is, where it makes an angle of more than $90^o$ with the director. As discussed in \cite{toner_wensink2016}, such turns are strongly suppressed if the ``energy barrier" $\Delta E$ against a reversal of the swimmer direction of motion (that is, the energy cost of the swimmer making an angle of $90^o$ with the local layer normal) is large compared to the thermal energy; i.e., if  $\Delta E\gg\kbt$. We have chosen our parameters to ensure this condition in our simulations.  Indeed, we have {\it never} observed a hairpin turn in our simulations. More importantly, we also expect that in many real experiments, deep within the smectic phase and for a strongly aligned swimmer, $\Delta E\gg\kbt$, so hairpins should be rare, if not non-existent,  as well.

We now proceed to analyze the implications of this theory for the motion of the swimmer.
We will start with the mean motion.
Taking the average of  \eq{align}, and recalling that $\langle\bff\rangle={\bf 0}$, we immediately obtain an expression for the mean position of the swimmer: 
\begin{eqnarray}
\langle\bfr_{s}(t) \rangle =v_st\langle\bn\rangle\equiv v_zt\bzh\,,
\label{meanrs}
\end{eqnarray}
where 
we have taken the mean direction of the layer normal $\bn$ to be along $\bzh$, and 
 the mean swimmer speed in the $z$ direction is
given by $v_z=v_s|\langle\bn\rangle|$.  
Thus, the mean motion of the swimmer is purely ballistic. Note that the speed $v_z$ of this motion is {\it not} $v_s$, due to the fact that fluctuations reduce $\langle\bn\rangle$ below $1$. Indeed,  the speed $v_z$ can {\it not} even be predicted by the continuum theory developed below, since the fluctuations which dominate this reduction
are predominantly short wavelength, and therefore not accurately described by the continuum, long-wavelength hydrodynamic  theory of smectics. 
%This domination by short wavelengths can be seen by noting that, roughly speaking, the mean squared fluctuations in Fourier space $\langle|\bfnp(\bfq)|^2\rangle$ of the components of the director perpendicular to $\bzh$ predicted by the Boltzmann weight associated with the Frank free energy \eq{Frank FE} obey $\langle|\bfnp(\bfq)|^2\rangle\propto1 /q^2 $. Since  the $\bfq$ space volume  in a spherical shell $q_0\le|\bfq \le 2q_0$ scales like $q_0^3$ in $d=3$, while the typical $\langle|\bfnp(\bfq)|^2\rangle$ in that shell scales like $1 / q_0^2$, the total contribution of such a shell to the mean squared real space fluctuations  $\langle|\bfnp(\bfr)|^2\rangle$, which contribution is proportional to $\int_{q_0\le|\bfq|\le2q_0} d^3 q \,  \left< |\bfnp \left(\bfq, t\right)|^2\right>$,  grows linearly with  $q_0$. That is, regions of larger $\bfq$ (i.e., smaller wavelength $1/ |\bfq| $) contribute more to $\langle|\bfnp(\bfr)|^2\rangle$ than regions of smaller $\bfq$. Hence, we can not compute these fluctuations from a long wavelength theory. We therefore cannot compute $\langle\bn\rangle$, and, therefore, cannot compute $v_z$. We must instead take it as yet another phenomenological parameter of our model. Equivalently, if we incorporate short wavelength effects by introducing an ultraviolet cutoff $\Lambda$ to our wavevector integrals, the value of $\langle\bn\rangle$, and, therefore, of $v_z$, will depend on $\Lambda$, which is another parameter. 
Nonetheless, we have still made a  universal scaling prediction: the mean motion of the swimmer is ballistic, as shown by  \eq{meanrs}.

We now turn to the fluctuations about this mean. Consider first  the mean squared lateral displacement of the swimmer:
\begin{eqnarray}
 \langle ( \Delta \bfr_{s}^{\perp}(t))^{2} \rangle \equiv \left<\left| \bfr_{s}^{\perp}(t) -
 \bfr_{s}^{\perp}(0) \right|^2 
\right> ,
\label{RW1}
\end{eqnarray}
{\it perpendicular} to the mean director of the smectic. Here
and throughout this paper, $\perp$ and $z$ denote directions
perpendicular to, and  along, the layer normal, respectively.

Using the projection of our equation of motion \eq{align} perpendicular to the mean layer normal direction $\bzh$, which reads
\beq
{d \bfr_{s}^{\perp}(t)\over dt}=v_s\bfnp(\bfr_{s},t)+\bff_{_\perp} \ ,
\label{alignperp}
\eeq
and integrating over time gives
\beq
\Delta\bfr_{s}^{\perp}(t)\equiv\bfr_{s}^{\perp}(t)-\bfr_{s}^{\perp}(0)=\int_0^tdt'\left(v_s\bfnp(\bfr_{s},t')+\bff_{_\perp}(t')\right) \,.
\label{delrperp}
\eeq
Squaring this, and averaging,  we find that $ \langle ( \Delta \bfr_{s}^{\perp}(t))^{2} \rangle$ is given by 
\begin{widetext}
\begin{eqnarray}
 \langle ( \Delta \bfr_{s}^{\perp}(t))^{2} \rangle =\int^t_0 d t^{\prime} \int^t_0 dt^{\prime\prime}
\left[ v_s^2\left<\bfnp(  \bfr_{s}(t^{\prime}),t^{\prime}) \cdot \bfnp(  \bfr_{s}
(t^{\prime\prime}), t^{\prime\prime})\right>+2v_s\left<\bfnp(  \bfr_{s}(t^{\prime}),t^{\prime}) \cdot  \bff_{_\perp}
(t^{\prime\prime})\right>+\left<\bff_{_\perp}
(t') \cdot \bff_{_\perp}
(t^{\prime\prime})\right>\right] \quad .
\label{RW 3}
\end{eqnarray}
\end{widetext}
 Using  the expressions \eq{ffcorr} and \eq{fncorr} for the two-point correlations of the Gaussian random velocity, we can immediately evaluate the last two terms,  denoted by $I_{2}$ and $I_{3}$, respectively. The first of them is
%\begin{eqnarray}
%I_2 &=&\int^t_0 d t^{\prime} \int^t_0 dt^{\prime\prime}
%2v_s\left<\bfnp(  \bfr_{s}(t^{\prime}),t^{\prime}) \cdot  \bff_{_\perp}
%(t^{\prime\prime})\right>=6\int^t_0 d t^{\prime} \int^t_0 dt^{\prime\prime}\Delta_{fn}
%\delta(t-t')= 6\Delta_{fn}t\quad,
%\label{nfcont1}
%\end{eqnarray}
\beq
I_2 =\int^t_0 d t^{\prime} \int^t_0 dt^{\prime\prime}
2v_s\left<\bfnp(  \bfr_{s}(t^{\prime}),t^{\prime}) \cdot  \bff_{_\perp}
(t^{\prime\prime})\right>= 6\Delta_{fn}t\quad ,
\label{nfcont1}
\eeq
while the second is
%\begin{eqnarray}
%I_3 &=&\int^t_0 d t^{\prime} \int^t_0 dt^{\prime\prime}
%\left<\bff_{_\perp}
%(t') \cdot \bff_{_\perp}
%(t^{\prime\prime})\right>=\int^t_0 d t^{\prime} \int^t_0 dt^{\prime\prime}\left[6\Delta_{I}
%+2\Delta_{A}
%|\bn_{\alpha}(\bfr_{s}(t),t)|^2\right]\delta(t-t')%\nonumber\\&
%= \left[6\Delta_{I}
%+2\Delta_{A}
%\right]t\quad.\nonumber\\
%\label{nfcont2}
%\end{eqnarray}
\beq
I_3 =\int^t_0 d t^{\prime} \int^t_0 dt^{\prime\prime}
\left<\bff_{_\perp}
(t') \cdot \bff_{_\perp}
(t^{\prime\prime})\right>= \left[6\Delta_{I}
+2\Delta_{A} 
\right]t ,
\label{nfcont2}
\eeq
Both of these terms are extremely boring: their contribution to 
the mean squared lateral wandering $ \langle ( \Delta \bfr_{s}^{\perp}(t))^{2} \rangle$ is simply conventionally diffusive: that is, proportional to time $t$. The anomalous diffusion that we predict comes entirely from the first term in  \eq{RW 3}:
%\begin{widetext}
\begin{eqnarray}
I_1 =v_s^2\int^t_0 d t^{\prime} \int^t_0 dt^{\prime\prime}
\left<\bfnp(  \bfr_{s}(t^{\prime}),t^{\prime}) \cdot \bfnp(  \bfr_{s}
(t^{\prime\prime}), t^{\prime\prime})\right> \,.
\nonumber\\
\label{RW 3.1}
\end{eqnarray}
%\end{widetext}
Because the smectic dynamics are invariant under space and time translations, the general director two point correlation function  depends only on the differences of the space and time coordinates; that is
%\begin{widetext}
\beq
C_\perp\equiv\left<\bfnp(  \bfr',t') \cdot \bfnp(  \bfr'', t^{\prime\prime})\right> = C_{_\perp} \left(\bfr' -
 \bfr'', t'- t^{ \prime \prime}\right) .
 \label{nemcorr}
\eeq
%\end{widetext}
Now in \eq{RW 3.1}, we need this correlation function evaluated when $\bfr'=\bfr_s(t')$ and $\bfr''=\bfr_s(t'')$. These vectors are given by:
%To evaluate  this expression , we need to know $\bfr_s(t)$. This is given by
\beq
\bfr_s (t) = \bfr_s (0) + v_{z}t \bzh +
\Delta\bfr_{s}^\perp(t)  \quad  .
\label{RW 5}
\eeq
%\end{widetext}
To proceed further, we need to calculate this director correlation, which is independent of the dynamics of the swimmer, but clearly does depend on the dynamics of the smectic. However, we will show, in the final subsection of this theoretical section, that in fact the motion of the swimmer is, at sufficiently long times, independent of the smectic dynamics, provided only that those dynamics do relax back to thermal equilibrium. 

However, that phrase "sufficiently long times" is highly loaded: for many real experimental systems, the time that must be reached before the asymptotic law for the lateral superdiffusion of the swimmer that we find below holds is astronomical. Fortunately, for shorter time scales, the behavior is still superdiffusive, but with a different scaling law. All of this will be discussed in the final subsection of this theoretical section.

In the next subsection, we will obtain the asymptotic superdiffusive scaling law that applies for all dynamical models. 

\subsection{Universal  asymptotic superdiffusion}

We will show in the next subsection that at very long times, the correlation function
 $C_{_\perp} \left(\bfr' -
 \bfr'', t'- t^{ \prime \prime}\right)$, when evaluated at typical values of  $\bfr'=\bfr_s(t')$ and $\bfr''=\bfr_s(t'')$, is well approximated by its equal time value; that is,
\beq
C_{_\perp} \left(\bfr_s(t') -
\bfr_s(t''), t'- t^{ \prime \prime}\right)\approx C_{_\perp} \left(\bfr_s(t') -
\bfr_s(t''), 0\right) .
 \label{etapprox}
 \eeq
However, we will also show in the next subsection that
for many realistic experimental systems, this "equal-time" approximation only holds for astronomically long time scales. For shorter time scales, we still predict superdiffusive behavior, but with a different scaling law. Our simulations, however, are done for a model and in a regime in which this asymptotic behavior is reached at quite short times.
For this section, we will simply assume that this equal time approximation (\ref{etapprox}) holds, and investigate its consequences for the lateral diffusion of the swimmer. 
 
The simplification provided by the equal time approximation is that equal time correlations can be calculated from equilibrium Boltzmann statistics. Indeed, many different dynamical models will relax back to the same equilibrium Boltzmann distribution, and, therefore, the same equal time correlation functions. This is particularly relevant for the simulations we perform here, since we simulated a model {\it without} momentum conservation, whereas any real bulk three dimensional smectic will, of course, have momentum conservation \footnote{The only exception to this would be smectics in aerogels. In these systems, the smectic can lose momentum to the aerogel mesh. However,. in this case the physics is quite different, due to the effects of the disorder of the aerogel on the smectic. See  L.\ Radzihovsky,  J.\ Toner,  and N.\ Clark, Science, {\bf 294}, 1074 
(2001), for a discussion of these systems}. We will not consider the motion of a swimmer in such systems here. Fortunately, since our model (by construction) relaxes back to the equilibrium state of a smectic A, it is guaranteed to have the same equal time correlation functions as the more realistic momentum conserving models that describe the experimentally relevant momentum conserving case. 

Having reduced our problem to the calculation of equilibrium,
equal-time correlations of the layer normal in a smectic A, we now proceed to calculate those correlations. 
This quite standard calculation starts with the observation that,
 in a smectic A, the layer normal $\bn(\bfr,t)$ is determined entirely by the smectic layer displacement field $u(\bfr,t)$ through the simple geometrical relation \cite{deGennes}:
\beq
\bn(\bfr,t)\approx\bhz-\nabla_\perp u(\bfr,t)  ,
\label{laynorm}
\eeq
where the approximate equality holds to linear order in $\nabla_\perp u(\bfr,t)$.
Fourier transforming this relation in space then implies
\begin{eqnarray}
 \left< |\bfnp \left(\bfq, t\right)|^2\right>=q_\perp^2 \left< |u \left(\bfq,t\right)|^2\right>  .
\label{nucorrq}
\end{eqnarray}
The calculation of the $u$-$u$ correlation function in this expression from equilibrium Boltzmann statistics requires only a knowledge of the equilibrium elastic Hamiltonian for layer positional fluctuations $u(\bfr)$. This is well known to be \cite{deGennes} \footnote{The full smectic elastic Hamiltonian contains anharmonic terms which are known to significantly change the scaling of the smectic correlations at long distances. However, the length scale at which these effects become important is also astronomically large in most real  systems. Hence we will ignore these effects here. The theory of these effects was worked out by  G. Grinstein and R. A. Pelcovits, 
{\it Phys. Rev. Lett.} {\bf 47},  856 (1981).}:
\begin{eqnarray}
H&=& \frac{1}{2}\int d^{3} r \left[B \left(\partial_z u\right)^2+ K  \left( \nabla^2_\perp u\right)^2\right]  .
\label{Elastic E}
\end{eqnarray}
From this model, it is straightforward to derive the required  $u$-$u$ correlation function in (\ref{nucorrq}) by Fourier transforming and applying equipartition. This gives the standard result \cite{deGennes}
\begin{eqnarray}
\left< |u \left(\bfq,t\right)|^2\right>={k_BT  \over G_\bfq}  ,
\label{usmeq}
\end{eqnarray}
where we have defined
\beq
G_\bfq\equiv
K q_{_\perp}^4+Bq_z^2 \ .
\label{Gdef}
\eeq
Using this in (\ref{nucorrq}) gives the director correlations in Fourier space
\begin{eqnarray}
 \left< |\bfnp \left(\bfq, t\right)|^2\right>={k_BT q_\perp^2 \over G_\bfq}  ,
\label{usmeq}
\end{eqnarray}
which in turn implies that real space director fluctuations are given by
%\begin{eqnarray}& C(\bfr, t)=\left< \bfnp \left(\bfr+{\bf R}, T+t\right)  \cdot \bfnp \left({\bf R}, T\right)\right> \nonumber \\ &=k_BT\int \frac{d^3 q}{8\pi^3} \,  {q_\perp^2e^{i\bfq\cdot \bfr-\Gamma G_\bfq t}\over G_\bfq}\quad .\label{ncorrft1}\end{eqnarray}
\begin{eqnarray}
C_{_\perp}(\bfr, 0) &=& \left< \bfnp \left(\bfr+{\bf R}, t\right)  \cdot \bfnp \left({\bf R}, t\right)\right> \nonumber \\
&=& k_BT\int \frac{d^3 q}{8\pi^3} \,  {q_\perp^2e^{i\bfq\cdot \bfr}\over G_\bfq}\quad .
\label{ncorret1}
\end{eqnarray}
Using this expression in our equal time approximation (\ref{etapprox}), and using that in turn in 
expression  (\ref{nemcorr}) for the nematic correlations and (\ref{RW 3.1}) for the superdiffusive integral $I_1$, we obtain
\beq
I_1 =v_s^2\int^t_0 d t^{\prime} \int^t_0 dt^{\prime\prime}
C\left(t^{\prime} -t^{\prime\prime}\right) .
\label{SD1}
\eeq
where we have defined
\beq 
C\left(\delta t\right)\equiv C_{_\perp} \left(\bfr_s(t+\delta t) -\bfr_s(t),  0\right)  ,
\label{Cdef}
\eeq
with $r_s\left(t\right)$ given by (\ref{RW 5}).
Using (\ref{ncorret1}) in this expression, and performing the integral over $q_z$ by
complex contour  techniques gives
\begin{eqnarray}
 C(\delta t)&=&\left< \bfnp \left(\Delta \bfr_{s}^{\perp}(\delta t)+v_s\delta t \bzh, 0\right)  \cdot \bfnp \left({\bf 0}, 0\right)\right> \nonumber \\ 
&=& k_BT\int \frac{d^2 q_{_\perp}}{8\pi^2\sqrt{BK}} \,  e^{i\bfqp\cdot\Delta \bfr_{s}^{\perp}(\delta t)-v_s \lambda q_{_\perp}^2 |\delta t|}\quad .
\label{ncorrft1}
\end{eqnarray}
where   $\lambda \equiv\sqrt{K / B}$ is the familiar smectic penetration length \cite{deGennes}, and we have defined $\Delta \bfr_{s}^{\perp}(\delta t)\equiv \bfr_s(t+\delta t)- \bfr_s(t)$. 
Doing the simple Gaussian integrals over the two components of $\bfqp$ gives:
\beq
C(\delta t) = \frac{ k_{B}T} { 8\pi v_{s} K|\delta t|} \langle \exp\left[-{\left( \Delta \bfr_{s}^{\perp}(\delta t)\right)^{2}\over 4\lambda v_s|\delta t|}\right] \rangle\,.
 \label{ncorrrealtsm}
\eeq
Noting that each of the two components $\Delta x$,  $\Delta y$ of $\Delta \bfr_{s}^{\perp}(\delta t)$ is a zero-mean Gaussian random variable (since it is the sum of Gaussian random variables, as it is linearly related to the noise, which is Gaussian, and the director $\bn$, whose fluctuations are also  Gaussian), and using the result for a zero mean Gaussian random variable $x$ that $ \langle \exp(-kx^{2}) \rangle=1 /\sqrt{1+2k \langle x^2 \rangle}$, we get 
\begin{eqnarray}
C(\delta t) &=& \frac{ k_{B}T} { 8\pi \left[v_{s} K|\delta t|+{2\pi K\left<\left( \Delta \bfr_{s}^{\perp}(\delta t)\right)^{2}\right>\over \lambda}\right]} \nonumber \\
& \approx & \frac{ k_{B}T} {2 \pi \sqrt{BK}\left<\left( \Delta \bfr_{s}^{\perp}(\delta t)\right)^{2}\right>},
 \label{ncorrrealtsm2}
\end{eqnarray}
where in the last, approximate, equality, we have assumed (as we will verify {\it a posteriori} is true in the limit $\delta t\rightarrow 0$) that $\left<\left( \Delta \bfr_{s}^{\perp}(\delta t)\right)^{2}\right>\gg \lambda v_{s} K|\delta t|$. This assumption amounts to assuming that there {\it is} anomalous diffusion in this case, as we will now show.

Using the approximate equality of (\ref{ncorrrealtsm2}) in (\ref{RW 3}), %and doing one of the time integrals, 
leads to a self-consistent equation for $\left<\left( \Delta \bfr_{s}^{\perp}(t)\right)^{2}\right>$:
\beq
\left<\left( \Delta \bfr_{s}^{\perp}(t)\right)^{2}\right>= \frac{ k_{B}T} {2 \pi\sqrt{BK} }\int^t_0 d t^{\prime}\int^t_0 d t^{\prime\prime}{1\over\left<\left( \Delta \bfr_{s}^{\perp}(t^{\prime}-t^{\prime\prime})\right)^{2}\right>}
\label{scons1}
\eeq

We will seek  a self-consistent solution to this equation of the form 
\beq
\left<\left( \Delta \bfr_{s}^{\perp}(t)\right)^{2}\right>= \overline{D}t\left[\ln\left({t\over t_0}\right)\right]^\alpha \ ,
\label{ansatz}
\eeq
where $t_0$ is a short-time cutoff, and  $\alpha$ and $ \overline{D}$ are, respectively, an exponent and a ``superdiffusion" constant, both of which  we will determine self-consistently. This leads to the condition:
\begin{widetext}
\beq
 \overline{D}t\left[\ln\left({t\over t_0}\right)\right]^\alpha= \frac{ k_{B}T} {2 \pi\overline{D}\sqrt{BK} }\int^t_0 d t^{\prime}\int^t_0 d t^{\prime\prime}{1\over |t^{\prime}-t^{\prime\prime}|\left[\ln\left({ |t^{\prime}-t^{\prime\prime}|'\over t_0}\right)\right]^\alpha}= \frac{ k_{B}T} {2 \pi\overline{D}\sqrt{BK}(1-\alpha) }t\left[\ln\left({t\over t_0}\right)\right]^{1-\alpha}\,.
\label{scons2}
\eeq
\end{widetext}
where in the second step the integrals over $t^\prime$ and $t^{\prime\prime}$ exclude the region in which $|t^\prime-t^{\prime\prime}|<t_0$. This expression
(\ref{scons2})
is clearly satisfied if  $\alpha=1-\alpha$, so $\alpha={1\over 2}$, and 
\beq
 \overline{D}={v_s\over(BK)^{1/4}}\sqrt{k_BT\over\pi} \ .
 \label{Dbar}
 \eeq
Inserting these results into our ansatz (\ref{ansatz}) leads to our final prediction for asymptotic lateral diffusion of the swimmer at asymptotically large times:
\beq
\left<\left( \Delta \bfr_{s}^{\perp}(t)\right)^{2}\right>= \overline{D}t\sqrt{\ln\left({t\over t_0}\right)} \ ,
\label{anomdifasy}
\eeq
with the anomalous diffusion constant $\overline{D}$ given by (\ref{Dbar}).

Recall that these results were derived on the {\it assumption} that the explicit dependence on $t^\prime-t^{\prime\prime}$ of $C_{_\perp}$ in (\ref{nemcorr}) could be neglected for sufficiently large times. In the next subsection, we will demonstrate that this is true both for the non-momentum conserving model that we simulate, and for real smectics, in which momentum is conserved. We will also estimate how large the time has to be before the asymptotic law (\ref{anomdifasy}) applies. This asymptotic time $t_a$ proves to be quite short for the non-momentum conserving model, but depends exponentially on parameters for momentum conserving models. As a result, for many momentum conserving models, $t_a$ is astronomically large, and a different, but still anomalous, scaling law, which we also derive below, applies.

\subsection{ Asymptotic time for different dynamical models}

There are a number of different dynamical models for $u(\bfr,t)$ that will relax to the equilibrium distribution for a smectic A. We simulate a simple, purely  relaxational model, which reduces at long wavelengths to:  
\beq
{\partial u(\bfr,t)\over \partial t} =-\Gamma {\delta H\over\delta u}+f(\bfr,t) \ ,
\label{EOMunoncon}
\eeq
\newline
where the Hamiltonian $H$ is given by (\ref{Elastic E}),
and  the noise $f$ in \eq{EOMunoncon} must obey the fluctuation-dissipation theorem, which implies: 
\begin{eqnarray}
   \langle f(\bfr,t)f(\bfr',t')\rangle=2\Gamma\kbt
\delta^3(\bfr-\bfr')\delta(t-t') \, .
\label{fcorrnoncon}
\end{eqnarray}
These dynamics differ from those of real bulk smectics \cite{Pershan} which are complicated by the coupling of the layer displacement field $u(\bfr,t)$ to background fluid flow. In the important range of wavevectors $q_z\lesssim\lambda q_\perp^2$, $|\bfq | \ll a^{-1}$ (where $a$ is the smectic layer spacing), these equations reduce to \cite{Pershan} 
\begin{eqnarray}
{\partial u(\bfr,t)\over \partial t} &=& g_z/\rho_0,  \\
\label{EOMucon}
{\partial g_z(\bfr,t)\over \partial t} &=& - {\delta H\over\delta u}+{\eta_2\over\rho_0}\nabla_\perp^2 g_z+f_z(\bfr,t),
\label{EOMvcon}
\end{eqnarray}
where $g_z$ is the local $z$-component of the momentum density of the smectic, $\rho_0$ is the mean density of the smectic, $\eta_2$ is one of the five viscosities characterizing the viscous response of uniaxial systems like smectics A, the elastic Hamiltonian $H$ is still given by (\ref{Elastic E}), and the noise $f_z$ is constrained by the fluctuation-dissipation theorem to satisfy 
\begin{eqnarray}
   \langle f(\bfr,t)f(\bfr',t')\rangle=2\eta_2\kbt\nabla_\perp^2
\delta^3(\bfr-\bfr')\delta(t-t') \, ,
\label{fcorrcon}
\end{eqnarray}
again in the wavevector regime of interest.

We will now consider each of these models in turn, and show that the long-time limit of the superdiffusive behavior is given by (\ref{anomdifasy}) for both of them. We will also calculate the asymptotic time $t_a$ for both models, and derive the alternative superdiffusive law for $t\ll t_a$ in the momentum conserving case, for which $t_a$ can be astronomically large.
%cons case.

\subsubsection{Non-momentum conserving model}

We seek the space and time dependent correlation function $C_{_\perp}(\bfr, t)$  in equation (\ref{nemcorr}) of the director. As before, we  will obtain this from the correlations of the displacement field $u(\bfr,t)$. These can readily be obtained from the equation of motion (\ref{EOMunoncon}) by Fourier transforming in space, solving the resultant ordinary differential equation for $u(\bfq, t)$ in terms of $f_z(\bfq,t)$, and autocorrelating the result at two different times. This gives 
\begin{eqnarray}
 \left< u \left(\bfq, t+\tau\right)u \left(-\bfq, \tau \right)\right>={k_BT  e^{-\Gamma G_\bfq t}
\over G_\bfq} \ .
\label{usmeq}
\end{eqnarray}
where we have defined
\beq
G_\bfq\equiv
K q_{_\perp}^4+Bq_z^2 \ .
\label{Gdef}
\eeq
This implies that director correlations in Fourier space are given by
\begin{eqnarray}
 \left< \bfnp \left(\bfq, t+\tau \right)\cdot\bfnp\left(-\bfq, \tau \right)\right>={k_BT q_\perp^2 e^{-\Gamma G_\bfq t}
\over G_\bfq} \ ,
\label{usmeq}
\end{eqnarray}
which in turn implies that real space director fluctuations are given by
%\begin{eqnarray}& C(\bfr, t)=\left< \bfnp \left(\bfr+{\bf R}, T+t\right)  \cdot \bfnp \left({\bf R}, T\right)\right> \nonumber \\ &=k_BT\int \frac{d^3 q}{8\pi^3} \,  {q_\perp^2e^{i\bfq\cdot \bfr-\Gamma G_\bfq t}\over G_\bfq}\quad .\label{ncorrft1}\end{eqnarray}
\begin{eqnarray}
C(\bfr, t) &=& \left< \bfnp \left(\bfr+{\bf R}, \tau+t\right)  \cdot \bfnp \left({\bf R}, \tau \right)\right>  \nonumber \\
&=& k_BT\int \frac{d^3 q}{8\pi^3} \,  {q_\perp^2e^{i\bfq\cdot \bfr-\Gamma G_\bfq t}\over G_\bfq}\quad .
\label{ncorrft2}
\end{eqnarray}
Changing variables of integration in this multiple integral from $\bfq_\perp$ to ${\bf Q}_\perp\equiv |\bfr_\perp |\bfq_\perp$ and from $q_z$ to $Q_z\equiv |\bfrp|^2q_z/\lambda$ enables us to rewrite this in a scaling form:
%\begin{widetext}
\beq
C(\bfr, t)={k_BT\over\sqrt{BK}|\bfr_\perp|^2}\Upsilon_N\left({\lambda z\over |\bfr_\perp|^2}, {\Gamma Kt\over|\bfr_\perp|^4}\right)\quad ,
\label{ncorrscale}
\eeq
%\end{widetext}
where we have defined the scaling function 
%\begin{widetext}
\beq
\Upsilon_N\left(\psi, \zeta\right) =\int \frac{d^3 Q}{8\pi^3}  \frac{Q_\perp^2\exp\left[i(Q_x+Q_z\psi)-\zeta (Q_z^2+Q_\perp^4)\right]}{Q_z^2+Q_\perp^4}
\,.
\label{ncorrft3}
\eeq
%\end{widetext}
To justify the equal-time approximation made in the preceding section, we need to show that the explicit time dependence of this correlation function (\ref{ncorrscale}) can be neglected. From the scaling form, we see that this will be a good approximation whenever the dimensionless  scaling variable $\zeta=
{\Gamma Kt\over|\bfrp|^4}$ associated with time is small; that is,  $\zeta\ll1$. Using our result   
(\ref{anomdifasy}) from the previous section for $\langle |\bfrp|^2 \rangle$, and estimating the typical value of $|\bfrp|^4\sim \langle |\bfrp|^2 \rangle^2$ leads to the condition for the validity of our asymptotic result (\ref{anomdifasy}):
 \beq
 {\Gamma K \over\Dbar^2 t\ln\left({t\over t_0}\right)}\ll1
 \label{tacond}
 \eeq
which is clearly always satisfied at long times. Indeed, it is satisfied whenever
\beq
t\gg t_a\equiv {\Gamma K \over\Dbar^2}= {\Gamma K\sqrt{BK} \over\kbt v_s^2} \ .
\label{tacon}
\eeq
Since $t_a$ is a fairly weak function  (i.e., algebraic, rather than exponential, as in the momentum conserving case) of the parameters of our model,  we expect it to be fairly easy to reach the asymptotic regime $t\gg t_a$, in which our asymptotic result (\ref{anomdifasy}) applies. And indeed, we find in our simulations that (\ref{anomdifasy}) holds from very early times out to the longest times we can simulate, as we have just predicted.

\subsubsection{Momentum conserving model}

In Fourier space, our momentum conserving model becomes
\begin{eqnarray}
{\partial u(\bfq,t)\over \partial t} &=& g_z(\bfq,t)/\rho_0 \ , \\
\label{EOMuconFT}
{\partial g_z(\bfq,t)\over \partial t} &=& - G_\bfq u-{\eta_2\over\rho_0}q_\perp^2 g_z(\bfq,t)+f_z(\bfq,t) \ ,
\label{EOMvconFT}
\end{eqnarray}
where $G_\bfq$ was defined in (\ref{Gdef}). If we assume we are in the Stokesian limit, in which the viscous ($\eta_2$) term dominates the inertial ${\partial g_z(\bfq,t)\over \partial t}$ term in (\ref{EOMvconFT}), then we can solve that equation directly for $g_z(\bfq,t)$, obtaining
\beq
g_z(\bfq,t)=-\rho_0\left({G_\bfq u(\bfq,t)-f_z(\bfq,t)\over\eta_2q_\perp^2}\right) \ .
\label{gzsol}
\eeq
Inserting this into the equation of motion (\ref{EOMuconFT}) for $u(\bfq,t)$ gives
\beq
{\partial u(\bfq,t)\over \partial t} =-\left({G_\bfq u(\bfq,t)-f_z(\bfq,t)\over\eta_2q_\perp^2}\right) \ .
\label{EOMustokes}
\eeq
We can now check {\it a posteriori} our assumption that we are in the Stokesian limit by using this expression to compute ${\partial g_z(\bfq,t)\over \partial t}$, and taking its ratio with the $\eta_2$ term in (\ref{EOMvconFT}). Doing so, and keeping only the $u$-dependent terms, we find this ratio is $R\equiv {\partial_t g_z(\bfq,t)\over {\eta_2\over\rho_0}q_\perp^2g_z}\sim{\rho_0G_\bfq\over\eta_2^2q_\perp^4}$. This ratio is clearly a monotonically increasing function of $q_z^2$, so it is biggest when $q_z=0$, at which point it is given by ${K\rho_0\over\eta_2^2}\equiv\chi$.  The smaller this ratio, the better our Stokesian approximation. For typical thermotropic smectics, $K\sim 5\times10^{-7}\rm{dynes}$, $\rho_0\sim1{\rm{gram}\over\rm{cm-sec}}$, and $\eta_2\sim 1 \rm{Poise}$, which gives $\chi\sim 5\times 10^{-7}$. Even in lyotropic smectics, for which $\eta_2$ approaches the viscosity of water, which is two orders of magnitude smaller than the value of $\eta_2$ we have just used, we still get $\chi\sim5\times 10^{-3}$, and this is before taking into account the reduction of $K$ due to dilution. So our Stokesian approximation is clearly a very good one in all cases.

The correlation functions of $u$ can now be obtained from (\ref{EOMustokes}) as in the nonconserving case, by Fourier transforming in space, solving the resultant ordinary differential equation for $u(\bfq, t)$ in terms of $f_z(\bfq,t)$, and autocorrelating the result at two different times. This gives
%For the more realistic momentum conserving case, it can be shown \cite{MRT} that 
\begin{eqnarray}
 \left< u \left(\bfq, t+\tau \right)u \left(-\bfq, \tau \right)\right>={k_BT  \exp\left(-{ G_\bfq\over\eta_2q_\perp^2}t\right)
\over G_\bfq} \ .
\label{usmeqwet}
\end{eqnarray}
%where $\eta_2$ is one of the viscosities characterizing the anisotropic viscous damping in a smectic A\cite{deGennes, MRT}.
This implies that director correlations in Fourier space are given by
\begin{eqnarray}
 \left< \bfnp \left(\bfq, t+\tau \right)\cdot\bfnp\left(-\bfq, \tau \right)\right>={k_BT q_\perp^2  \exp\left(-{ G_\bfq\over\eta_2q_\perp^2}t\right)
\over G_\bfq} \ ,\nonumber\\
\label{nsmeqwet}
\end{eqnarray}
which in turn implies that real space director fluctuations are given by
%\begin{eqnarray}& C(\bfr, t)=\left< \bfnp \left(\bfr+{\bf R}, T+t\right)  \cdot \bfnp \left({\bf R}, T\right)\right> \nonumber \\ &=k_BT\int \frac{d^3 q}{8\pi^3} \,  {q_\perp^2e^{i\bfq\cdot \bfr-\Gamma G_\bfq t}\over G_\bfq}\quad .\label{ncorrft1}\end{eqnarray}
\begin{eqnarray}
C(\bfr, t) &=& \left< \bfnp \left(\bfr+{\bf R}, \tau+t\right)  \cdot \bfnp \left({\bf R}, \tau \right)\right> \nonumber \\
 &=& k_BT\int \frac{d^3 q}{8\pi^3} \,  {q_\perp^2e^{i\bfq\cdot \bfr-{ G_\bfq\over\eta_2q_\perp^2}t}\over G_\bfq}\quad.
\label{ncorrrscon}
\end{eqnarray}
Changing variables of integration in this multiple integral from $\bfq_\perp$ to ${\bf Q}_\perp\equiv |\bfr_\perp |\bfq_\perp$ and from $q_z$ to $Q_z\equiv |\bfrp|^2q_z/\lambda$ enables us to rewrite this in a scaling form:
%\begin{widetext}
\beq
C(\bfr, t)={k_BT\over\sqrt{BK}|\bfr_\perp|^2}\Upsilon_C\left({\lambda z\over |\bfr_\perp|^2}, {Kt\over\eta_2|\bfr_\perp|^2}\right)\quad ,
\label{concorrscale}
\eeq
%\end{widetext}
where we have defined the scaling function 
\beq
\Upsilon_C\left(\psi, \zeta\right) =\int \frac{d^3 Q}{8\pi^3}\frac{Q_\perp^2\exp\left[i(Q_x+Q_z\psi)-\zeta \left(\frac{Q_{z}^{2}}{Q_{\perp}^{2}}+Q_\perp^4\right)\right]} {Q_z^2+Q_\perp^4} .
\label{ncorrft4}
\eeq
To justify the equal-time approximation made in the preceding section, we need to show that the explicit time dependence of this correlation function (\ref{ncorrscale}) can be neglected. From the scaling form, we see that this will be a good approximation whenever the dimensionless  scaling variable $\zeta=
{Kt\over\eta_2|\bfr_\perp|^2}$ associated with time is small; that is,  $\zeta\ll1$. Using our result   
(\ref{anomdifasy}) from the previous section for $\langle |\bfrp|^2 \rangle $, and estimating the typical value of $|\bfrp|^2\sim \langle |\bfrp|^2 \rangle$ leads to the condition for the validity of our asymptotic result (\ref{anomdifasy}):
 \beq
 {K t\over\eta_2\Dbar t\sqrt{\ln\left({t\over t_0}\right)}}\ll1
 \label{tacondcon}
 \eeq
which is clearly always satisfied at long times. Indeed, it is satisfied whenever
\beq
t\gg t_a\equiv t_0\exp\left[\left({K\over\eta_2\Dbar}\right)^2\right]\equiv t_0\exp\left[\left({v_c \over v_s}\right)^2\right] \  ,
\label{tacon}
\eeq
where we have defined a characteristic velocity
\beq
v_c\equiv{B^{1/4}K^{5/4}\over\eta_2}\sqrt{\pi\over\kbt} \ .
\label{vcdef}
\eeq
In deriving this expression, we have used equation (\ref{Dbar}) for $\Dbar$.

We see from equation (\ref{tacon}) that, in contrast to the non-momentum conserving case, when momentum is conserved, the asymptotic time (\ref{tacon}) is extremely sensitive (indeed, exponentially so) to material parameters, and to the speed of the swimmer. It can also become astronomically large. Taking typical numbers for a thermotropic smectic, such as $B\sim 5\times10^{7}{\rm{dynes}\over\rm{cm}^2}$ , $K\sim 5\times10^{-7}\rm{dynes}$, and $\eta_2\sim 1 \rm{Poise}$ gives $v_c\sim 5{\rm{cm}\over\rm{sec}}$. Putting a bacterium with a swimming speed of $v_s\sim50{\mu\over\rm{sec}}$ in such a smectic, we see that equation (\ref{tacon}) implies an asymptotic time of $t_a=t_0\exp(10^4)$, which, for any reasonable $t_0$, is far longer than the age of the universe!

However, the extreme exponential sensitivity of the asymptotic time $t_a$ means that it should be achievable in other systems. Lyotropic smectics are a good candidate. In a highly dilute lyotropic smectic, the elastic constants obey \cite{Helfrich} $B\sim{\left(\kbt\right)^2\over\kappa\ell^3}$,  and $K\sim{\kappa\over\ell}$, where $\ell$ is the lamellar spacing and $\kappa$ is the bend rigidity per unit area of a single lamella. Inserting these expressions into our expression (\ref{vcdef}) gives
\beq
v_c\sim{\kappa\over\eta_2\ell^2} \ .
\label{vclam}
\eeq
Thus, for very dilute systems, in which $\ell$ is large, we can make
the characteristic speed $v_c$ very small. Taking a typical lamellar bend stiffness $\kappa\sim5\times10^{-14}\rm{ergs}$ and noting that for a highly dilute lamellar phase, we expect and $\eta_2\sim\eta_{_{H_2O}}\sim10^{-2} \rm{Poise}$, 
we obtain from (\ref{vclam})
\beq
v_c\sim5{\rm{\mu}\over\rm{sec}}\left({1\mu\over\ell}\right)^2 \ ,
\label{vclamnum}
\eeq
which implies from (\ref{tacon}) that for a lamellar phase with a layer spacing of $\ell = 0.3$  micron, and a bacteria swimming at $v_s=50{\mu\over\rm{sec}}$, the asymptotic time would be
\beq
t_a\sim t_0e^{(10/9)^2}\approx3.44t_0 \ ,
\label{tanum}
\eeq
which should be quite experimentally accessible.

What about those cases in which $t_a$ is astronomically large? We can show that in those cases, there is also anomalous diffusion, but with a different scaling law. To see this, we note that $t\ll t_a$, the time variable is now the dominant one in the scaling law (\ref{concorrscale}) for the correlation function. Hence, we can evaluate that correlation function setting $\bfrp$ and $z$ to zero in our general expression (\ref{ncorrrscon}) for $C_{_\perp}$.
This gives
\begin{eqnarray}
C(\bfr={\bf 0}, t) &=& \left< \bfnp \left(\bfr+{\bf R}, \tau+t\right)  \cdot \bfnp \left({\bf R}, \tau \right)\right>  \nonumber \\
& = & k_BT\int \frac{d^3 q}{8\pi^3} \,  {q_\perp^2e^{-{ G_\bfq\over\eta_2q_\perp^2}t}\over G_\bfq}\quad .
\label{ncorrtcon}
\end{eqnarray}
With the change of variables $\bfq_\perp$ to ${\bf Q}_\perp\equiv \sqrt{Kt\over\eta_2}\bfq_\perp$ and from $q_z$ to $Q_z\equiv q_z{\lambda Kt\over\eta_2}$, we can pull the time dependence out of this expression, obtaining
%\begin{widetext}
\beq
C(\bfr={\bf 0}, t)=\left({\kbt\eta_2\over8\pi^3\sqrt{BK^3}}\right){f(1)\over t}\ ,
\label{ncorrtcon2}
\eeq
where we have defined
\beq
f(x)\equiv\int d^3Qq  {Q_\perp^2e^{-x\left({ Q_z^2+Q_\perp^4\over Q_\perp^2}\right)}\over Q_z^2+Q_\perp^4}\quad .
\label{fdef}
\eeq
This function can easily be evaluated by differentiating it with respect to $x$; this gives
\beq
f^\prime(x)=\int d^3Q {e^{-x{ Q_z^2+Q_\perp^4\over Q_\perp^2}}}\quad .
\label{fp}
\eeq
Performing the Gaussian integral over $Q_z$ gives
\beq
f^\prime(x)=\int d^2Q_\perp e^{-xQ_\perp^2}Q_\perp\sqrt{\pi\over x}\quad .
\label{fp2}
\eeq
The integral $d^2Q_\perp$ is also elementary; we thereby obtain
\beq
f^\prime(x)=-{\pi^2\over2 x^2}\quad .
\label{fp3}
\eeq
Integrating this, and determining the unknown constant of integration by noting, from inspection of (\ref{fp}), that $f(x\to\infty)\to 0$, we obtain 
\beq
f(x)={\pi^2\over2 x}\quad .
\label{f2}
\eeq
Using this in (\ref{ncorrtcon2}) gives
\beq
C(\bfr={\bf 0}, t)=\left({\kbt\eta_2\over16\pi\sqrt{BK^3}}\right){1\over t}\ .
\label{ncorrtcon3}
\eeq
Now replacing $\left<\bfnp(  \bfr_{s}(t^{\prime}),t^{\prime}) \cdot \bfnp(  \bfr_{s}
(t^{\prime\prime}), t^{\prime\prime})\right>$ in equation (\ref{RW 3}) with $C(\bfr={\bf 0}, t^\prime-t^{\prime\prime})$ from this expression, and doing the $t^\prime$ and $t^{\prime\prime}$ integrals in (\ref{RW 3}) gives, again, anomalous diffusion, but with a different scaling law:
\beq
 \langle ( \Delta \bfr_{s}^{\perp}(t))^{2} \rangle=\left({\kbt\eta_2\over16\pi\sqrt{BK^3}}\right)t\ln\left({t\over t_0}\right) \ .
 \label{t<ta}
 \eeq
This is the expression that is most experimentally relevant to thermotropic smectics, or to lyotropics at lower dilutions (i.e., lamellar spacings $\ell\ll1\mu$).

\section{Simulation model}

\begin{figure*} 
\begin{center} 
\includegraphics[width= 2 \columnwidth]{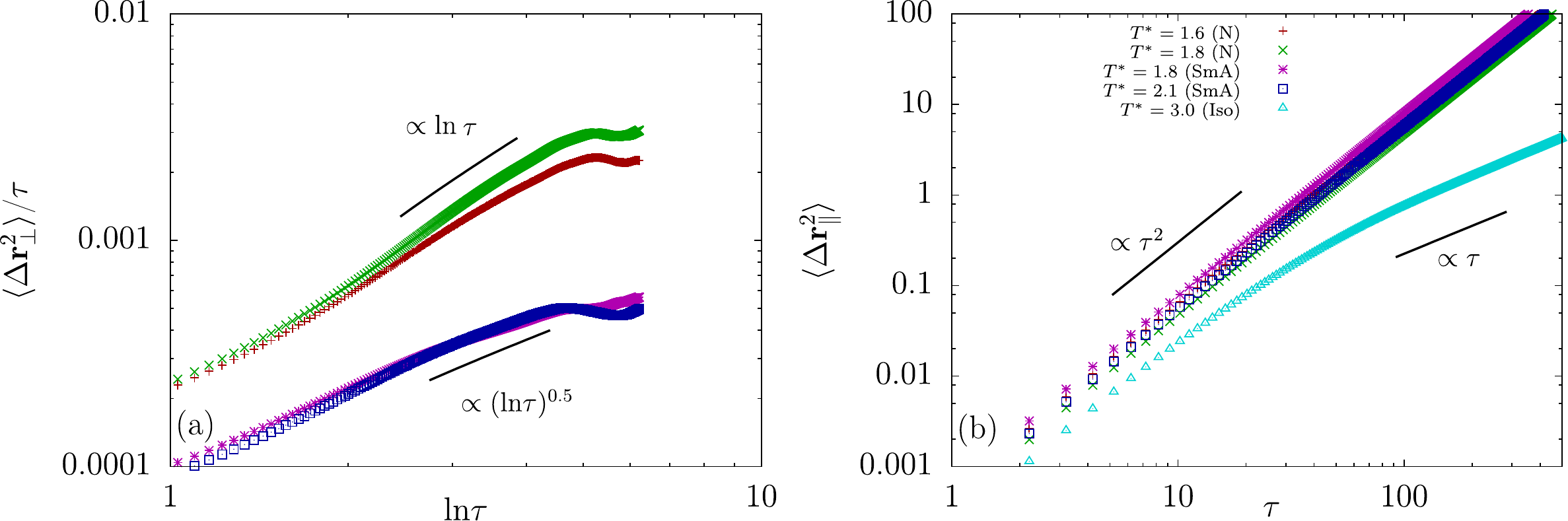} 
\caption{ \label{fig1} 
MSDs of a swimmer moving through a nematic and smectic medium. Shown is the
contribution (a) perpendicular and (b) parallel to the layer normal at various temperatures $T^{\ast}$.  The rod densities corresponding to the various phases are $\rho ^{\ast} = 0.22$  (Iso), $\rho ^{\ast}  = 0.3$ (N) and $\rho ^{\ast}  = 0.5$ (SmA).  The long-time dynamics transverse to the director is
anomalous and exhibits a distinct long-time logarithmic behavior with
scaling exponent $\alpha \approx 1$ for the nematic and $\alpha \approx 0.5$
for the smectic phase. For the isotropic phase, standard long-time swimmer diffusion $\langle \Delta {\bf r}^{2} \rangle \propto \tau$ is observed, as it should.} 
\end{center} 
\end{figure*}

We will now attempt to corroborate the theoretical predictions for the long-time swimmer dynamics in the smectic fluid using a particle-based simulation model. 
The model potential employed in our simulations is designed to generate stable nematic and smectic (A) phases at low temperature while producing trivial isotropic fluids at high temperature.  It corresponds to a simple soft-core potential proposed in Ref.\ \cite{jlintuvuori2008} and models the interaction energy $U_{sc}$ between two soft spherocylinders at centre-of-mass  displacement $\Delta \br$ with orientation unit vectors $\bhu$ and $\bhu^{\prime}$:
\begin{equation} \label{juho}
U_{sc}= \left\{
  \begin{array}{l l}
    u_{m} (1-\sigma)^2+\epsilon, &  \sigma < 1\\
     u_{m} (1-\sigma)^2- U_{a}(1-\sigma)^4 +\epsilon, & 1 \leq \sigma < \sigma_{c}\\
     0 &  \sigma \geq \sigma_{c}
  \end{array} \right. 
\end{equation}
where  $\sigma(\Delta \br , \bhu, \bhu^{\prime})$ denotes the shortest distance between two short spherocylinders of length $L$  and diameter $\sigma_{0}$ at fixed mutual orientation. The attractive part of the potential takes the form 
\beq
U_{a} = u_{a} -   5 \epsilon_{1} {\mathcal P}_{2}(\bhu \cdot \bhu^{\prime}) -   5 \epsilon_{2}  [{\mathcal P}_{2}(\Delta \hat{\br} \cdot \bhu) + {\mathcal P}_{2} (\Delta \hat{\br} \cdot \bhu^{\prime} )  ] 
\label{ua}
\eeq
in  terms of a second-order Legendre polynomial ${\mathcal P}_{2}$ and centre-of-mass distance unit vector $\Delta \hat{\br}$ . Furthermore, $\epsilon (\Delta \hat{\br} , \bhu, \bhu^{\prime} ) = -u_{m}^{2}/4U_{a}(\Delta \hat{\br}, \bhu, \bhu^{\prime})$ is the maximum well depth for the configuration chosen such as to guarantee the potential and its first derivative to reach zero at the cut-off distance $\sigma_{c} (\Delta \hat{\br} , \bhu, \bhu^{\prime} )= 1 + \sqrt{u_{m}/2U_{a}(\Delta \hat{\br}, \bhu, \bhu^{\prime})} $. The shape of the soft-core potential can be judiciously tuned through the four-parameter combination $\{ u_{m}, u_{a}, \epsilon_{1}, \epsilon_{2}\}$ enabling facile simulation of a range of liquid crystalline mesophases \cite{jlintuvuori2008}.

The swimmer is described as a point particle with position $\br_{s}$ and  orientation unit vector $\bhu_{s}$. It interacts with the surrounding soft rods by means of  a coupling potential \cite{toner_wensink2016} of strength $\epsilon_s$:
\begin{equation}
U_{s}= \epsilon_{s} \sum_i {\mathcal P}_{2}(\bhu_{i} \cdot \bhu_{s} ) g( r_{i,s}) 
\label{p2couple}
\end{equation}
The coupling potential decays with increasing distance  $r_{i,s}=||\br_{i,s}||$  between the swimmer and the soft rods through a Gaussian $g( r_{i,s})=\exp[- (r_{i,s}/\sigma_{s})^2]$ with $\sigma_{s}$ a characteristic length scale setting the coupling range. 

The microscopic equations of motion for the positional coordinates describe overdamped Brownian motion of each rod $i$:
\begin{eqnarray}
\bxi \cdot \partial_{t} \br_{i} &=&  -\nabla_{\br_{i}} U( \{ \br_{i} , \bhu_{i} \} )+ {\bf \bar{f}}_{i} 
\end{eqnarray}
where $\bxi  = \xi^{\parallel} \bhu \bhu + \xi^{\perp} ({\bf I} - \bhu \bhu) $ denotes the translational friction tensor of a uniaxial rod. The first term on the rhs is a direct force on rod $i$  due to presence of neighboring liquid crystal particles via the total potential energy $U = \frac{1}{2} \sum_{i,j} U_{sc}$, assumed pairwise additive. Furthermore,  $\bar{f}^{\alpha} = \sqrt{2 \xi_{\alpha} k_{B} T }R_{\alpha}(t)$ is a Gaussian random force acting on each rod with zero mean $\langle R_{\alpha}(t) \rangle=0$ and variance $\langle R_{\alpha}(t) R_{\gamma}(t^{\prime}) \rangle=  \delta_{\alpha \gamma}\delta (t -t^{\prime})$ with $\alpha, \gamma$ indicating the components of a rod-based orthonormal frame ${\bf \bar{f}}  = \bar{f}^{\parallel} \bhu + \bar{f}^{\perp1} \bhe_{1} +  \bar{f}^{\perp2} \bhe_{2}$.
The equation of motion for the orientation of the soft-core particles follows from a similar balance of torques, via
\begin{eqnarray}
\xi_{R} \partial_{t} \bhu_{i} =  (\bw_{i} + \bw_{i,s} + \bar{\bw}) \times \bhu_{i}
\end{eqnarray}
with $\xi_{R}$ the rotational friction factor, $\bw_{i}  = \lambda_{i}\bhu_{i}  - \partial U/\partial \bhu_{i}$ the torque on particle $i$ due to the surrounding rods (with $\lambda_{i}$ a Lagrange multiplier ensuring normalization of $\bhu_{i}$), and $\bw_{i,s} =\lambda_{i}\bhu_{i} - \partial   U_{s} / \partial \bhu_{i}$ the contribution imparted by the presence of the swimmer, and $\bar{\omega}_{\alpha} = \sqrt{2 \xi_{R} k_{B} T} R_{\alpha}(t) $ 
a random Gaussian torque within the orthonormal particle frame so that $\alpha = \{  \perp1, \perp2  \}$.  The  geometric factors
$\{ \xi_{\parallel}, \xi_{\perp}, \xi_{R} \}$ depend solely on the rod aspect ratio
$p=(L + \sigma_{0})/\sigma_{0}>0$, and we adopt  the standard expressions for rod-like macromolecules, as given in \cite{tirado}:
\begin{eqnarray}
\frac{\xi_{0}}{\xi_{\perp}}= \frac{1}{4 \pi } \left( \ln{p} + 0.839 + 0.185/p + 0.233/p^2 \right), \nonumber  \\
\frac{\xi_{0}}{\xi_{\parallel}} = \frac{1}{2 \pi } \left( \ln{p} - 0.207 + 0.980/p - 0.133/p^2 \right), \nonumber  \\
\frac{\xi_{R0}}{\xi_{R}} = \frac{3}{2 \pi p^2 } \left( \ln{p} - 0.207 + 0.980/p - 0.133/p^2 \right)
\label{xis} 
\end{eqnarray}
with $\xi_{0} $ and $\xi_{R0}$ the friction factors of a reference sphere with radius $\sigma_{0}$. Defining $D_{0} = k_{B}T /\xi_{0}$ as the translational diffusion coefficient of a reference sphere. 

The trajectory $\{ \br_{s}(t), \bhu_{s}(t) \} $ of the swimmer is governed by the following equations of motion:
\begin{eqnarray}
\bxi_{s}  \cdot \partial_{t} \br_{s} &=&  f_{a} \bhu_{s} \nonumber \\
 \xi_{Rs} \partial_{t} \bhu_{s} &=&  (\bw_{s,i} + \bar{\bw}_{s}) \times \bhu_{s}
\end{eqnarray}
with $\xi_{Rs} $ the rotational friction factor and $\bxi_{s} $ the translational friction tensor of the swimmer, $\bw_{s,i} = \lambda_{s} \bhu_{s} - \partial U_{s}/ \partial \bhu_{s} $ the torque acting on the swimmer imparted by neighboring soft rods and $ \bar{\bw}_{s}$ a random torque. The geometric factors are identical to \eq{xis} for a given hydrodynamic swimmer aspect ratio $p_{s} > 1$ which may be different from that of the soft rods $p$.     The translational noise on the swimmer is ignored as it is assumed to be negligible compared to the orientational noise (as is the case for motile bacteria \cite{2011DrescherEtAl}).  
Throughout this study, we use dimensionless  expressions for time  $\tau=t D_{0}/\sigma_0^2$, temperature  $T^{*}=k_BT/u_{a}$, particle density $\rho^{*} = N\sigma_{0}^{3}/V$ (with $N$ the number of particles and $V$ the system volume) and  active force $f_{a}^*=f_a \sigma_{0} /\epsilon_{s}$.

The interaction parameters (in units $k_{B}T$) for the current model are the following; $u_{m}=25$, $ u_{a}=150$, $\epsilon_1=10$, $\epsilon_2=-2$. The rod aspect ratio is fixed at $p=L/\sigma_{0}=3.0$.  For these values the system undergoes a transformation from an isotropic, to a nematic and a smectic phase upon lowering the temperature and/or increasing particle density. We employ a cubic simulation box with periodic boundary conditions in all three directions. The initial configuration represents $N=3500$ rods forming a square smectic lattice  at a given density $\rho^*$. It is allowed to melt during an equilibration run of at least $\Delta \tau=2000$. Once the host phase is equilibrated, a swimmer is placed in the centre of the simulation box and the whole system is re-equilibrated during at least $\Delta \tau=2000$.  To avoid numerical artifacts the time step associated with the linear discretization of the equations of motion is kept sufficiently small ($\delta \tau< 0.001$). The parameters defining the interaction between  the swimmer and the host rods are;  $\epsilon_s=1 k_{B}T$ , $\sigma_s=(L+\sigma_{0})/2$ for the coupling strength and range, respectively, and  $p_{s}=5$ for the swimmer aspect ratio. Production runs during which observables of interest were recorded span a total time interval of at least $\Delta \tau=5000$.  

Let us define  the displacement vectors $\Delta  \bfr^{\parallel}_{s}(\tau)\equiv ((\br_{s} (\tau) - \br_{s}(0)) \cdot \bn) \bn $ along, and $\Delta  \bfr^{\perp}_{s}(\tau)\equiv  \Delta \bfr_{s} (\tau) - \Delta\bfr^{\parallel}_{s}(\tau)$ perpendicular to, the layer normal $\bn$,  with $\Delta \bfr_{s}(\tau) = \bfr_{s}(\tau)  - \bfr_{s}(0)$. We then determined from our simulations the
mean squared displacements  $\langle ( \Delta  \bfr^{\parallel}_{s}(\tau) )^{2} \rangle$ and $\langle ( \Delta  \bfr^{\perp}_{s}(\tau) )^{2} \rangle$, where $\langle \cdots \rangle$ denotes a time-average in the steady state.
We  also monitor the mean-squared rotation of the swimmer via:
 \beq
 \langle   {\mathcal P}_{n} ( \bhu_{s}(\tau) \cdot \bhu_{s}(0)) \rangle, \hspace{0.5cm} n=1,2
\eeq
These quantities enable us to gauge the typical  rotational relaxation time of the swimmer in relation to the symmetry of the medium and temperature. Since the  effective shape of the swimmer is strongly anisotropic, i.e.  $p_{s} >1$, the rate of hairpin turns is extremely small \cite{toner_wensink2016} and no such events are recorded during the course of our simulations except, of course, in the case of an isotropic background medium (at $T^{\ast} \gg 1$) where the orientational fluctuations of the swimmer are strong and  the long-time dynamics becomes strictly diffusive.  

\begin{figure*} 
\begin{center} 
\includegraphics[width= 1.8 \columnwidth]{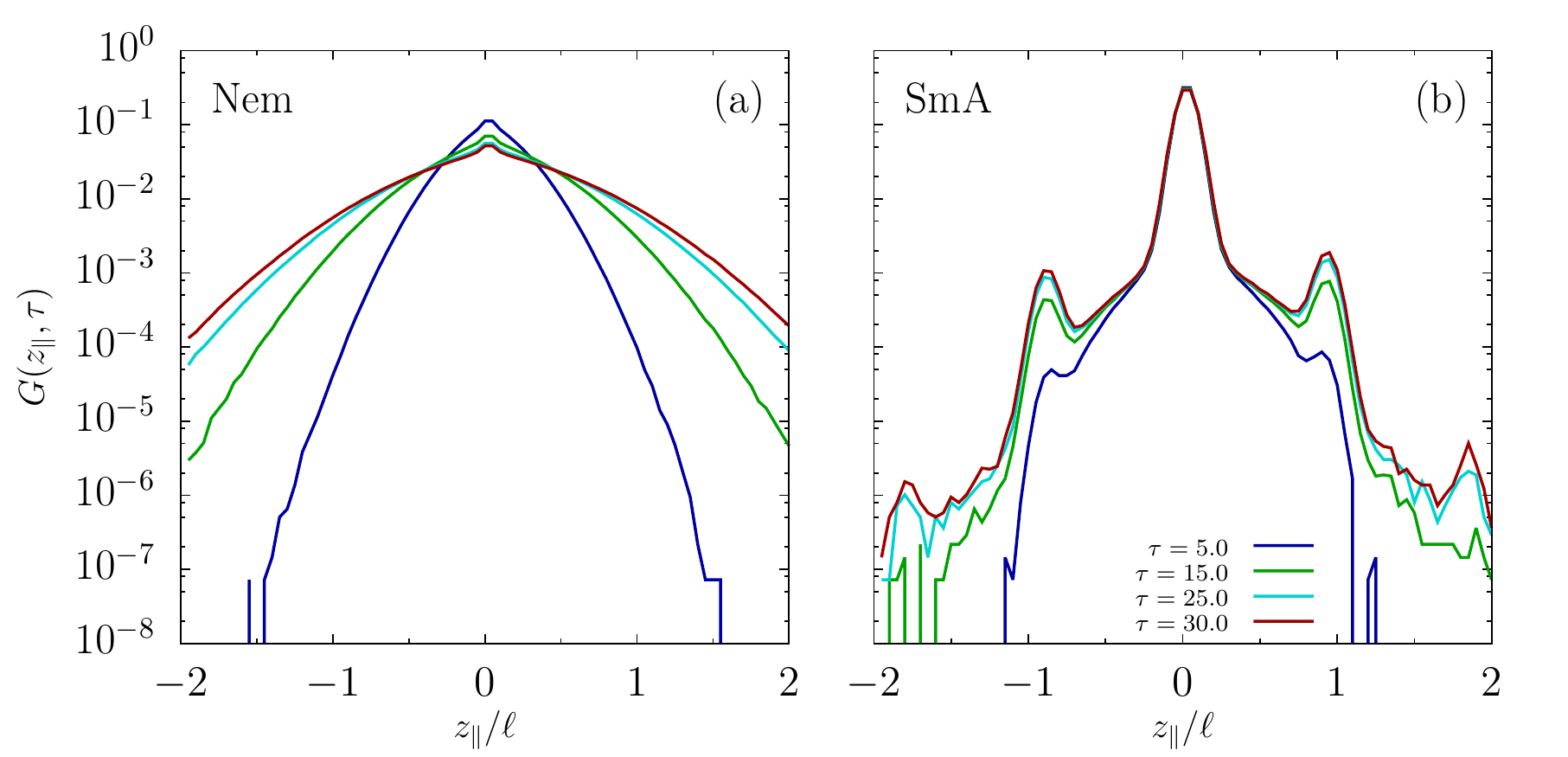} 
\caption{ \label{fig2} Van Hove correlation function $G(z_{\parallel}, \tau) $ for the host rods forming a (a) nematic phase ($T^{\ast} =1.8 $, $\rho^{\ast} = 0.3$) and (b) smectic phase ($T^{\ast} =1.8 $, $\rho^{\ast} = 0.5$ ). Curves are depicted at various time $\tau$. The distinct shoulders at discrete rod lengths $\ell$ indicate the diffusive barriers imparted by the lamellar microstructure of the smectic phase.    } 
\end{center} 
\end{figure*}

An overview of the transverse MSDs for a swimmer moving is shown in \fig{fig1}a.   In this particular representation,  the logarithmic long-time tails clearly show up in a linear fashion with the scaling exponent for the smectic systems differing significantly from those of the nematic systems.  Note that, in Fig. 1a,   time is represented on an extremely compressed $\ln (\ln \tau)$-scale and deviations from linear scaling occur at very long times  where the statistics is no longer fully reliable. Running systematically longer simulation runs will obviously remedy this.   The fitted exponents ($\alpha = 0.5 \pm 0.1$ for the smectic and $\alpha = 1.0 \pm 0.1$ for the nematic) are in agreement with the ones established  from our theoretical model.  As expected, the mean-squared displacement parallel to the director remains ballistic throughout the sampled time interval for the nematic and smectic phases (suggesting hairpin turns to be completely absent), but for the isotropic phase it crosses over  from ballistic to diffusive $\propto \tau$ beyond the typical effective rotation time of the swimmer. The periodic boundary conditions employed in the simulation impart a weak bias on the layer normal which usually  points along either of the Cartesian axes of the simulation box, but it may also point along the box diagonal as we observe for the smectic phase. The effect of this bias can be systematically weakened by increasing the system size.  We find, however, that the latter has no measurable impact on the long-time scaling of the transverse MSD implying that the simulation set-up does not break the rotational invariance of the nematic or smectic fluid.  

The distinct lamellar signature of the smectic phase can be probed through the self-part of the van Hove function:
\beq
G(z_{\parallel}, \tau) =  \frac{1}{N} \left  \langle  \sum_{i=1}^{N} \int d r_{\parallel} \delta (z_{\parallel} - |\Delta  \bfr_{i}^{\parallel}(\tau) | ) \right \rangle 
\eeq
representing  the average distribution of displacements of the host particles along the layer normal over time. The results, depicted  in \fig{fig2}, enable a clear distinction between the Gaussian patterns of a spatially uniform nematic fluid,  whereas the shoulders point to a lamellar microstructure of the smectic host phase.

In our simulations we have also tested an alternative model in which a swimmer is conceived simply by rendering one of the passive host rods active by applying an active force $f_{a}$ along its main orientation axis. In this model, the coupling between the swimmer and the host rods proceeds via the short-ranged, {\em direct} interaction given by \eq{juho}. It turns out that both point-${\mathcal P}_{2}$-type swimmer and rod-based swimmer  produce the same asymptotic scaling of the transverse (and parallel) MSDs. This illustrates that the long-time scaling of the swimmer superdiffusion  does not depend on the details of the swimmer-host coupling provided its  propulsion direction is aligned {\em along} the local nematic director.  A clear advantage of the rod-based swimmer is that enables us to unambiguously recover standard long-time diffusion in an isotropic fluid (see Fig. 1b), while the mean-field ${\mathcal P}_{2}$ coupling interaction \eq{p2couple} yields spurious results for these dilute environments.    

\section{Conclusions}

In this paper we have addressed the long-term motility of a swimmer dispersed in a smectic A (or lamellar) fluid phase.   Using hydrodynamic theory for the swimmer dynamics subject to a fluctuating smectic director field,  we have derived universal scaling expressions for the mean-squared swimmer displacement perpendicular to the director. The predictions are relevant to a vast range of smectic phases of both thermotropic and lyotropic origin, and are independent of the dynamical process through which the nematic director field relaxes towards equilibrium.  We find that the long-time lateral diffusion of the swimmer  across the smectic membranes exhibits a logarithmic [$\propto t\sqrt{ \ln (t/t_{0})}$] scaling with time $t$ which is distinctly different from the anomalous swimmer diffusion [$\propto t \ln (t/t_{0})$]  in nematic fluids \cite{toner_wensink2016}  or the trivial long-time diffusion ($\propto t$) encountered in isotropic media.   We corroborate  our predictions using particle-based simulation of an active point-particle moving through a smectic A phase composed of soft mesogens.  Upon increasing temperature (and reducing particle concentration),  the smectic host transforms into a nematic and, subsequently, an  isotropic fluid. The concomitant transversal mean-squared swimmer displacement  is fundamentally different in each of these phases and the measured long-time scaling laws are in full agreement with theory.  Given the universality of our theoretical predictions one could envisage as a possible experimental application  measuring the long-time diffusion of swimming micro-organisms or active colloids propelling though liquid-crystalline media as a dynamical probe  to identifying the microstructure of the anisotropic host phase. Efforts to relate our theoretical findings to experimental model systems of external field-controlled active colloids moving through thermotropic liquid crystals  \cite{sagues2013, Sagues} are currently being undertaken.  

\acknowledgements{We thank J. Ign\'{e}s-Mullol and F. Sagu\'{e}s for helpful discussions. C.F.C. and H.H.W.  gratefully acknowledge funding from the French  National Research Agency (ANR) through the {\em ANR-JCJC} grant ``UPSCALE".  H. L. was supported by the DFG within project LO 418/20-1. J.T. thanks the Institut f\"{u}r Theoretische Physik II: Weiche Materie, Heinrich-Heine-Universit\"{a}t, D\"{u}sseldorf; the Max Planck Institute for the Physics of Complex Systems  Dresden; the Department of Bioengineering at Imperial College, London; The Higgs Centre for Theoretical Physics at the University of Edinburgh; and the Lorentz Center of Leiden University, for their hospitality while this work was underway.

\bibliographystyle{apsrev4-1}
\bibliography{refs}

\end{document}